\documentstyle[aps,psfig,twocolumn]{revtex}
 
\begin{document}

\title{Phase ordering of
two-dimensional symmetric binary fluids: a droplet scaling state}
\author{Alexander J. Wagner and M. E. Cates\\
Department of Physics and Astronomy, University of Edinburgh,\\ 
JCMB Kings Buildings, Mayfield Road, Edinburgh EH9 3JZ, U.K.}

\maketitle
\begin{abstract}
The late-stage phase ordering, in $d=2$ dimensions, of symmetric fluid
mixtures violates dynamical scaling. We show however that, even at
50/50 volume fractions, if an asymmetric droplet morphology is
initially present then this sustains itself, throughout the viscous
hydrodynamic regime, by a `coalescence-induced coalescence'
mechanism. Scaling is recovered (with length scale $l \sim t$, as in
$d=3$). The crossover to the inertial hydrodynamic regime is delayed
even longer than in $d=3$; on entering it, full symmetry is finally
restored and we find $l\sim t^{2/3}$, regardless of the initial
state. \end{abstract}

When a symmetric binary fluid is quenched below its critical
temperature it will phase separate into domains. This initial stage
of phase separation is diffusive; but in this letter we are concerned
with the hydrodynamic stage of the process, which follows
afterwards. The evolution equations can be taken \cite{bray} as
the Navier-Stokes and the advective diffusion equations:
\begin{eqnarray}
\rho D_t {\bf u} &=& - \nabla P 
- {\phi} \nabla \mu_\phi + \eta \nabla^2 {\bf u} \label{ns}\\
D_t\phi &=& \nabla \left(M\nabla \mu_\phi\right) \label{ad}
\end{eqnarray}
Here $\bf u$ is the fluid velocity, $D_t$ the material derivative, $P$
pressure, $\mu_\phi$ the chemical potential for the order parameter
$\phi$, $\eta$ the viscosity, $\rho$ the density and $M$ an Onsager
coefficient. As usual, $\mu_\phi = \delta F/\delta \phi$ with
$F[\phi]$ a (symmetric) free energy functional; once the interfaces
are sharp, the source term ($\sim \phi\nabla\mu_\phi$) in Eq. \ref{ns}
is just the Laplace pressure of interfacial curvature \cite{bray}.

These equations are nonlinear, but deterministic; thermal noise should
be irrelevant at late times. The only source of noise is then in the
initial condition, and it is often assumed that the physics is the
same for all conditions without long-range correlations
\cite{bray,kendon,bhattacharya}.  But we find below that the nonlinear
dynamics are more subtle: for two dimensional 50/50 binary fluid
mixtures of equal viscosity, the hydrodynamic coarsening depends very
strongly on the topology \cite{footeuler} of the sharp interfaces
formed by the early time (short length-scale) physics, previously
assumed irrelevant. Specifically, droplet morphologies evolve quite
unlike bicontinuous ones: dynamical scaling, absent for bicontinuous
states in $d=2$ dimensions \cite{prl}, is recovered for a new,
self-sustaining droplet state.

Such scaling requires that any averaged property of the system is
invariant if all lengths are scaled by a single length scale
$L(t)$. Whereas at early times diffusive phase-ordering leads to
$L\propto t^{1/3}$ in both $d=2$ and $d=3$, scaling of interfacial
curvatures with $L$ implies that at later times the chemical potential
obeys $\mu_\phi \sim \sigma/L(t)$ where $\sigma$ is the surface
tension (computable from $F[\phi]$).  Dimensional analysis of Eqs.1,2
then identifies two subsequent regimes, the viscous hydrodynamic (VH
\cite{siggia}) and the inertial hydrodynamic (IH \cite{oldfurukawa})
in which the driving force is balanced by viscosity and inertia
respectively \cite{kendon2}.

There is good numerical evidence for this behavior in bicontinuous
$d=3$ mixtures \cite{kendon}: allowing for a nonuniversal early time
diffusive offset, and choosing the reduced unit of time as $t_0 \equiv
\eta^3/\rho\sigma^2=1$, one finds a universal scaling curve $l(t)$
where $l(t) = L(t)\rho\sigma/\eta^2$ is the reduced coarsening
length. Choosing $L \equiv 2\pi \int S(k) dk/\int k S(k) dk$ with
$S(k)$ the density autocorrelator, one then finds that the viscous and
inertial asymptotes are $l=0.07 t$ and $l=0.9 t^{2/3}$; these cross at
$t^* \simeq 10^4$, a large value (though formally still $O(1)$)
\cite{kendon}.

In two dimensions, in contrast, the VH regime was recently found to
violate scaling altogether \cite{prl,furukawa}. Starting from a 50/50
bicontinuous state, a cascade of small droplets is left behind by the
large connected domains as they coarsen; these droplets, which must
coalesce in order to grow, evolve more slowly than the larger domains
around them. Length measures defined ({\em e.g.}) from different
moments of the structure factor do not scale.  Furukawa\cite{furukawa}
found, however, that the size of the {\it largest} domains does scale
as $l\sim t$ throughout VH; and that, in the IH regime, full scaling
is recovered with $l\sim t^{2/3}$.

For symmetric binary fluids, droplet morphologies usually arise for an
asymmetric composition. Until recently these morphologies were held to
give non-hydrodynamic phase ordering by either the evaporation-condensation
mechanism or the {\it Brownian} coalescence of droplets, each of which
has $t^{1/3}$ growth \cite{bray}. But Tanaka \cite{tanaka} discovered
that such droplet states can coarsen via a `coalescence induced
coalescence' (CIC) mechanism; Nicolayev {\it et
al.}\cite{nicolayev} considered this for some simplified $d=3$
geometries, and predicted the recovery of a VH
($l\sim t$) growth regime mediated by CIC for $\phi > 0.26$ in $d=3$.

In $d=2$, droplet morphologies can, however, also arise in symmetric (50/50)
quenches whenever there is asymmetry in the dynamics. This occurs,
{\em e.g.}, when the two components have different diffusivities so
that $M$ is strongly $\phi$-dependent \cite{ahluwalia}; or when one
phase is viscoelastic \cite{veps}. In the first case, the evolution
equation should approach that of a fully symmetric mixture on entering
VH.  In the second case (since the mean shear rate is of order $\dot
l/l \sim 1/t$), one must wait until $t$ exceeds the Maxwell time; this
could still be early in VH \cite{footvisco}. In either case the
dynamic asymmetry is then forgotten, {\it except} that it has
instilled local correlations in the ``initial'' condition for
subsequent events.  It is usual to argue that these correlations
cannot influence the late time scaling states once the domain scale
$L(t)$ is larger than the original correlation length; indeed,
Bhattacharya {\it et al.}\cite{bhattacharya} thereby argued that
viscoelastic phase separation could not lead to new scaling states. We
now show that, in contrast to such expectations, a distinct
CIC droplet scaling state does exist for a 50/50 quench, and survives
throughout the VH regime.

We use a well-established lattice Boltzmann method\cite{orlandini} to
simulate Eqs. 1,2. We initialize each simulation with a droplet
morphology, at 50/50 volume fraction, generated from a white-noise
initial state using a separate viscoelastic lattice Boltzmann model
\cite{veps}. (The initial droplets are at rest.) Thereafter the
evolution equations are fully symmetric: the two fluids are Newtonian
and have equal mobilities and viscosities. We choose parameters such
that the phase ordering is dominantly hydrodynamic, ensuring that
residual diffusion is negligible at the $<1$\% level \cite{kendon}. The
algorithm shows good isotropy for the shape of an equilibrium droplet
\cite{yeomPRE54}, but some dynamic anisotropy is noticeable in the IH
regime (visible in Figure 2).  Reduced physical units $l,t$ are defined
as above, and nonuniversal (diffusive) time offsets for each simulation
run calculated \cite{kendon}. All the simulations shown here were
performed on a $1024^2$ lattice, but terminated at $L \le 256$ to
mininize finite size artifacts \cite{kendon}. The surface tension is
$10^{-2}$, the density is 2, and we varied the viscosity from 0.5 to
0.001 (all in lattice units).

We used three different length-scale 
measures\cite{prl,shear}, derived from the number of domains $N$, the
interface length $L_I$ and the derivatives of the order-parameter $\phi$:
\begin{equation}
L^\# = \sqrt{\frac{\Lambda_x \Lambda_y}{N}} \;\;
L^\circ = \sqrt{2 \pi} \frac{\Lambda_x \Lambda_y}{L_I}\;\;
L^\partial =  \frac{2\sum_{\bf x}\phi^2}
	{\mbox{\rm Tr}\left(\sum_{\bf x} \partial_{\bf
	x} \phi \partial_{\bf x}\phi\right) }
\end{equation}
where $\lambda$ is the average of the eigenvalues of the derivative
matrix $\partial_{\bf x} \phi \partial_{\bf x} \phi$, and
$\Lambda_{x,y}$ are the sample dimensions.  The lengths $L^\#, L^\circ$
and $L^\partial$ are normalized such that each gives the unit cell
spacing for a square lattice of droplets of 50\% volume fraction.
For bicontinuous morphologies in $d=3$, all are 
comparable to $L(t)$ as defined previously; for example, $L^\partial
\simeq 1.2 L$.


We show in Fig. 1 the evolution of the droplet morphology for a
simulation run that lies wholly within the VH regime. After a brief
transient (leading to the first coalescence) the coarsening mechanism
is CIC, as first seen in experiments by Tanaka \cite{tanaka}. The
results are reminiscent of the predictions for CIC in $d=3$
\cite{nicolayev}, where it was predicted that the droplets that just
underwent a coalescence would be the next to coalesce. Instead we
observe that in $d=2$ the coalescence of two droplets usually induces
a coalescence elsewhere. (This is especially clear in animations
\cite{href}.) Perhaps for this reason, the CIC process does not (in
$d=2$) lead rapidly to percolation of the droplets as was suggested
for $d=3$ \cite{nicolayev}. Such percolation would cause a rapid
crossover to bicontinuity.  Instead, the system achieves a
self-sustaining droplet state that, in contrast to the bicontinuous
analogue \cite{prl}, shows good dynamical scaling (Fig.1). This
behavior is seen for all runs that lie within the VH regime as
identified from the universal scaling plot found below (Fig.3).

\begin{figure}
\begin{center}
\begin{minipage}{8cm}
\begin{minipage}{3.5cm}
\centerline{\psfig{figure=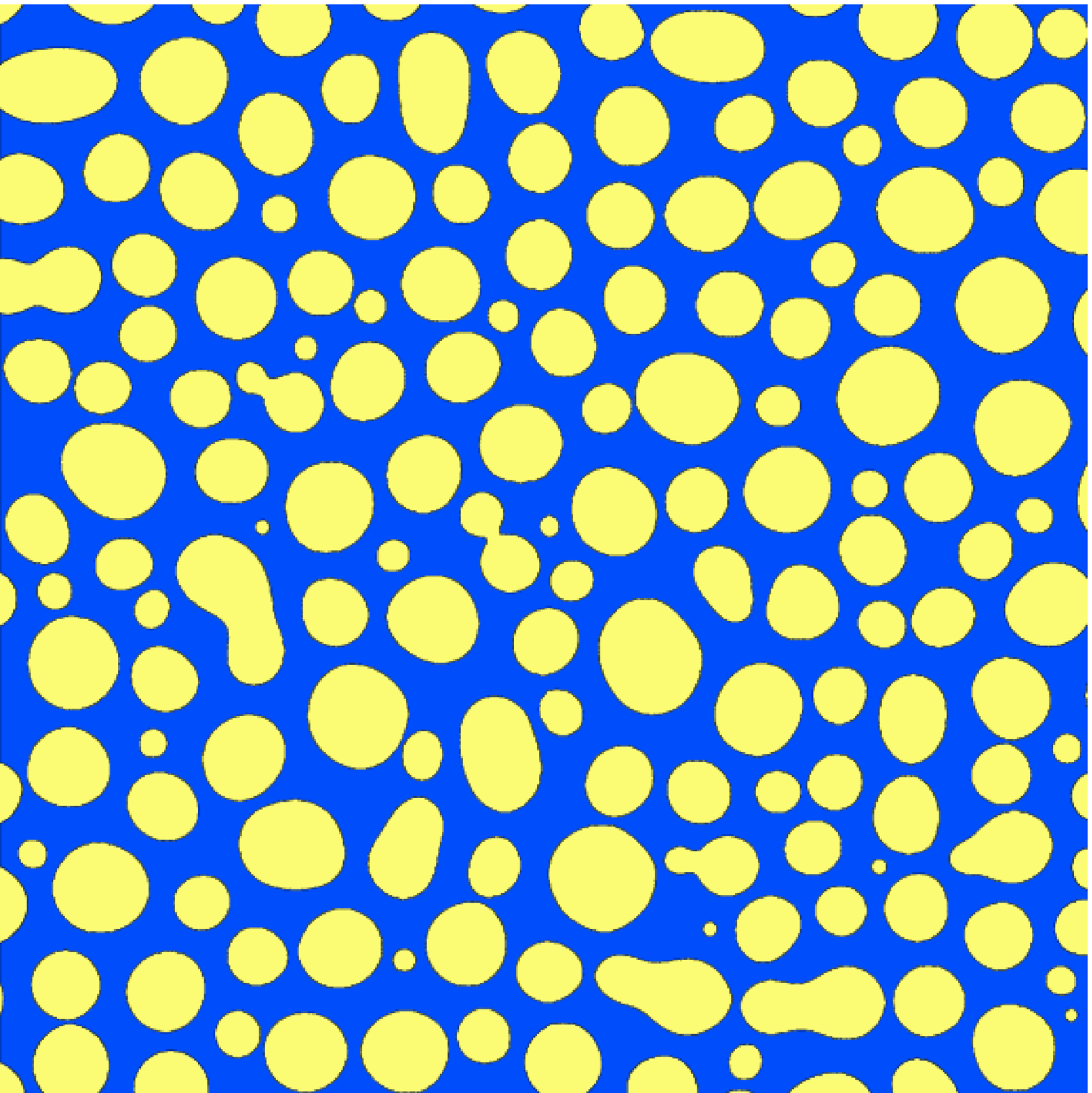,width=3cm}}
\end{minipage}
\hspace{0.3cm}
\begin{minipage}{3.5cm}
\centerline{\psfig{figure=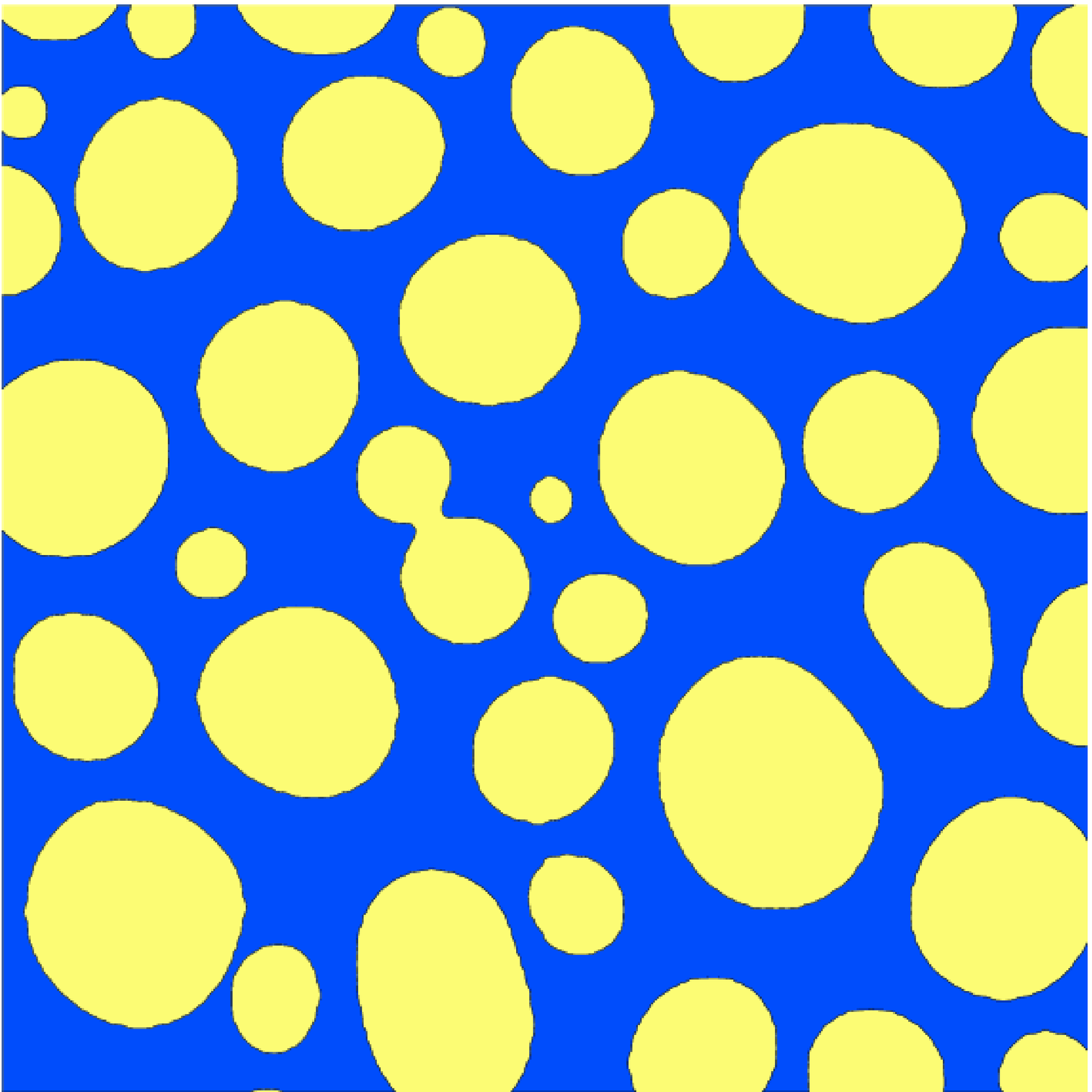,width=3cm}}
\end{minipage}\\ \vspace{0.1cm}

\begin{minipage}{3.5cm}
\centerline{\psfig{figure=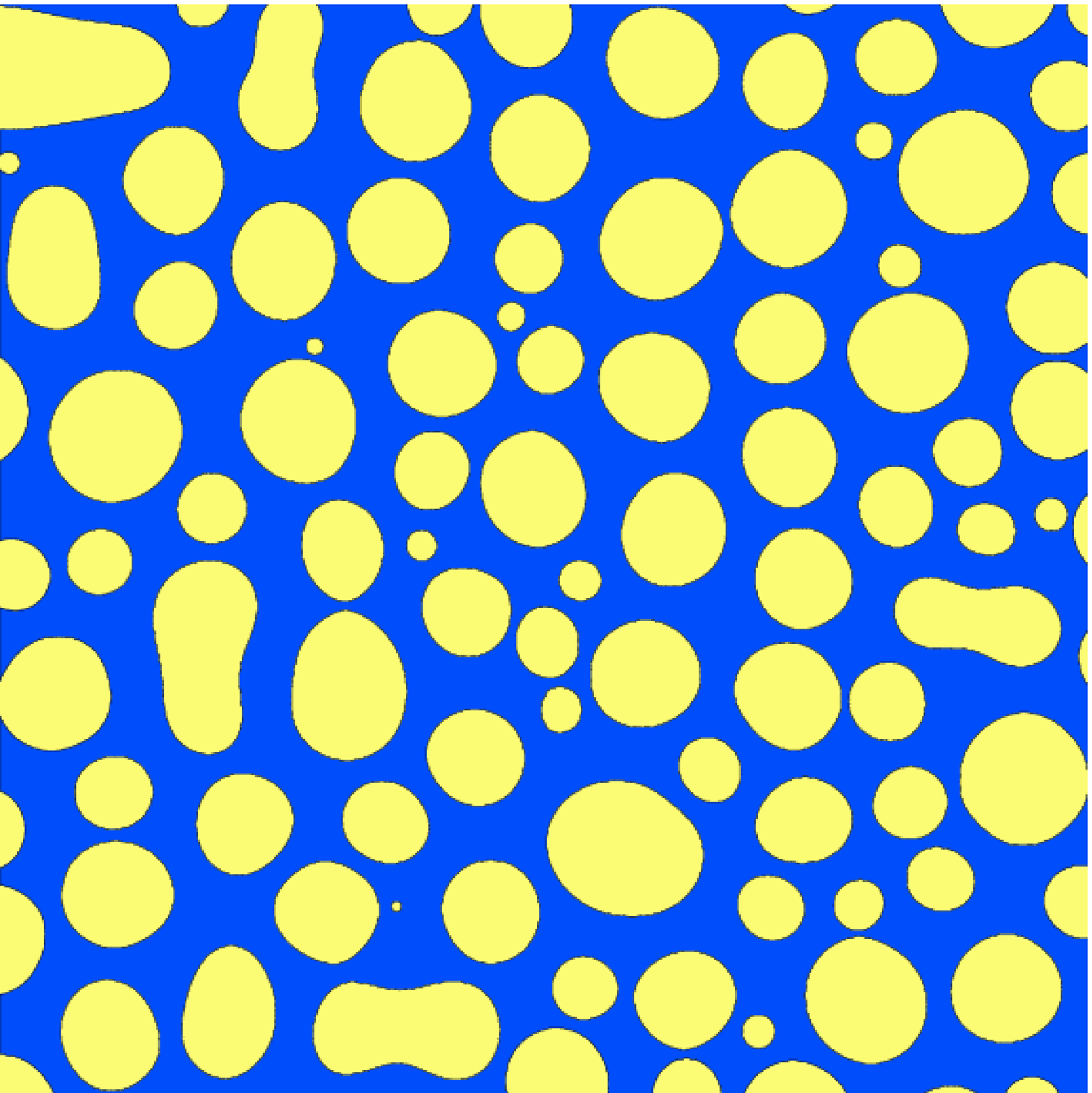,width=3cm}}
\end{minipage}
\hspace{0.3cm}
\begin{minipage}{3.5cm}
\centerline{\psfig{figure=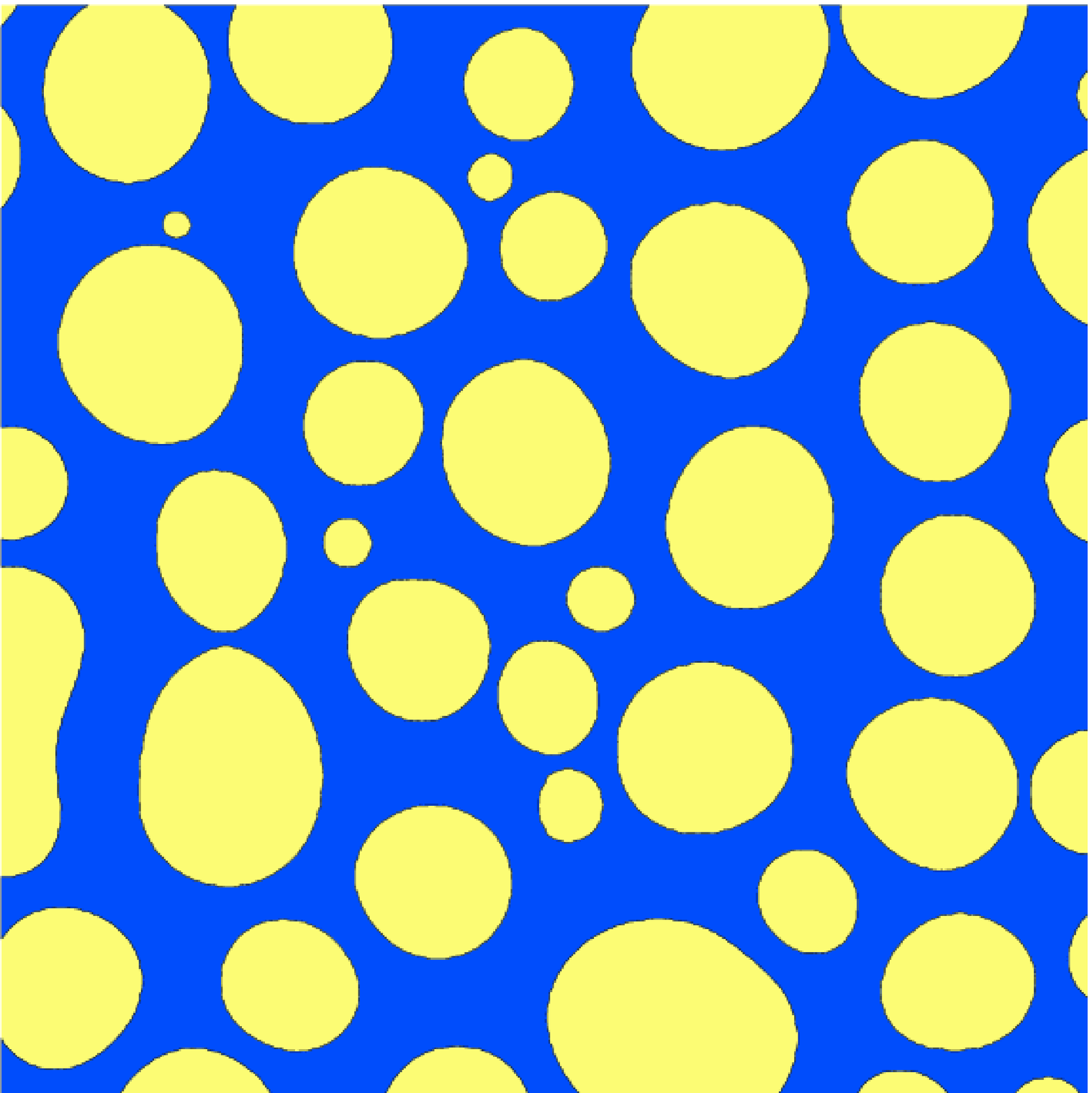,width=3cm}}
\end{minipage}\\ \vspace{0.1cm}

\begin{minipage}{3.5cm}
\centerline{\psfig{figure=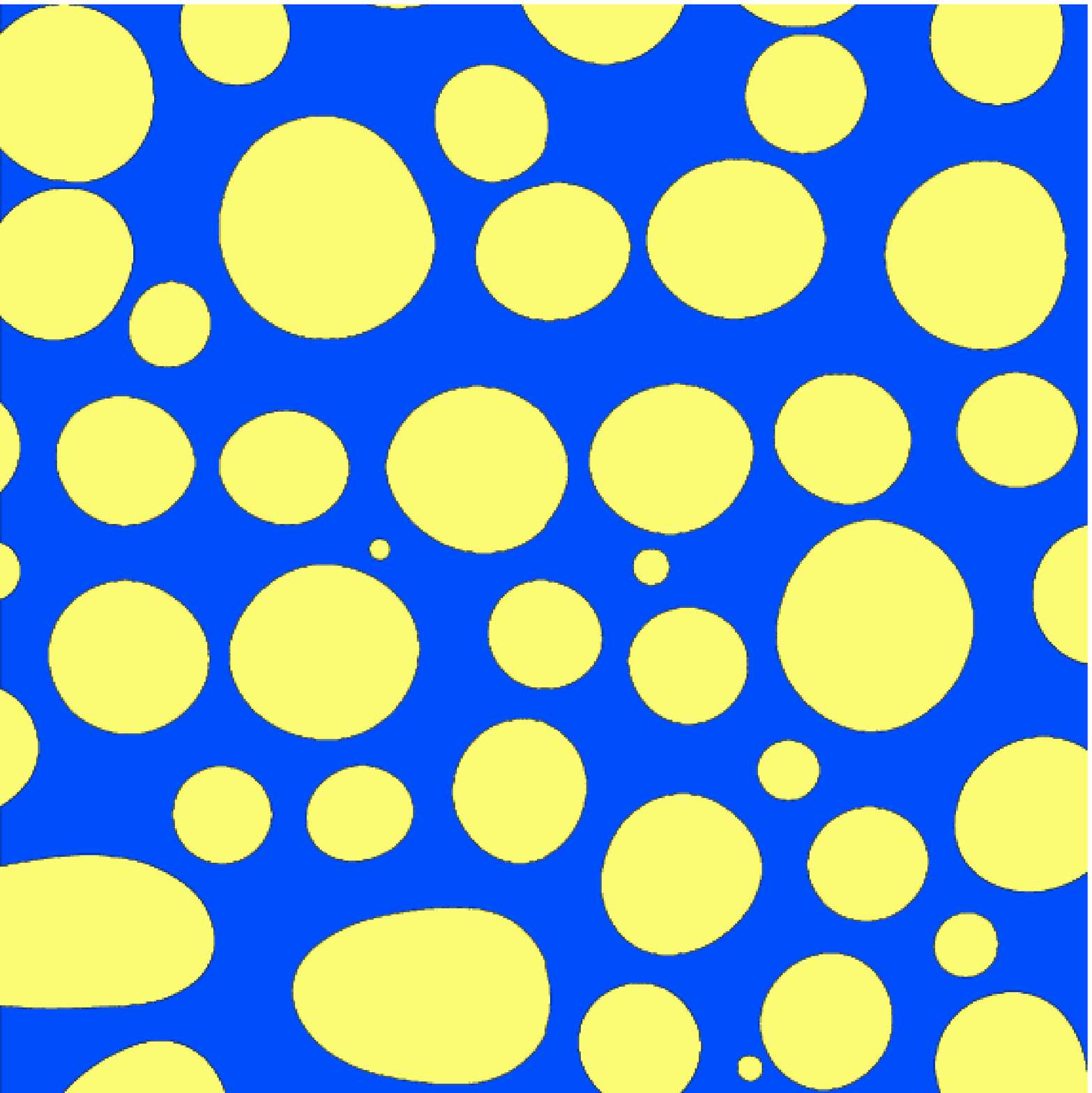,width=3cm}}
\end{minipage}
\hspace{0.3cm}
\begin{minipage}{3.5cm}
\centerline{\psfig{figure=snu0.11_86481.eps,width=3cm}}
\end{minipage}
\begin{center}
(a) \hspace{2.3cm} (b) 
\end{center}
\begin{minipage}{8cm}
\centerline{\psfig{figure=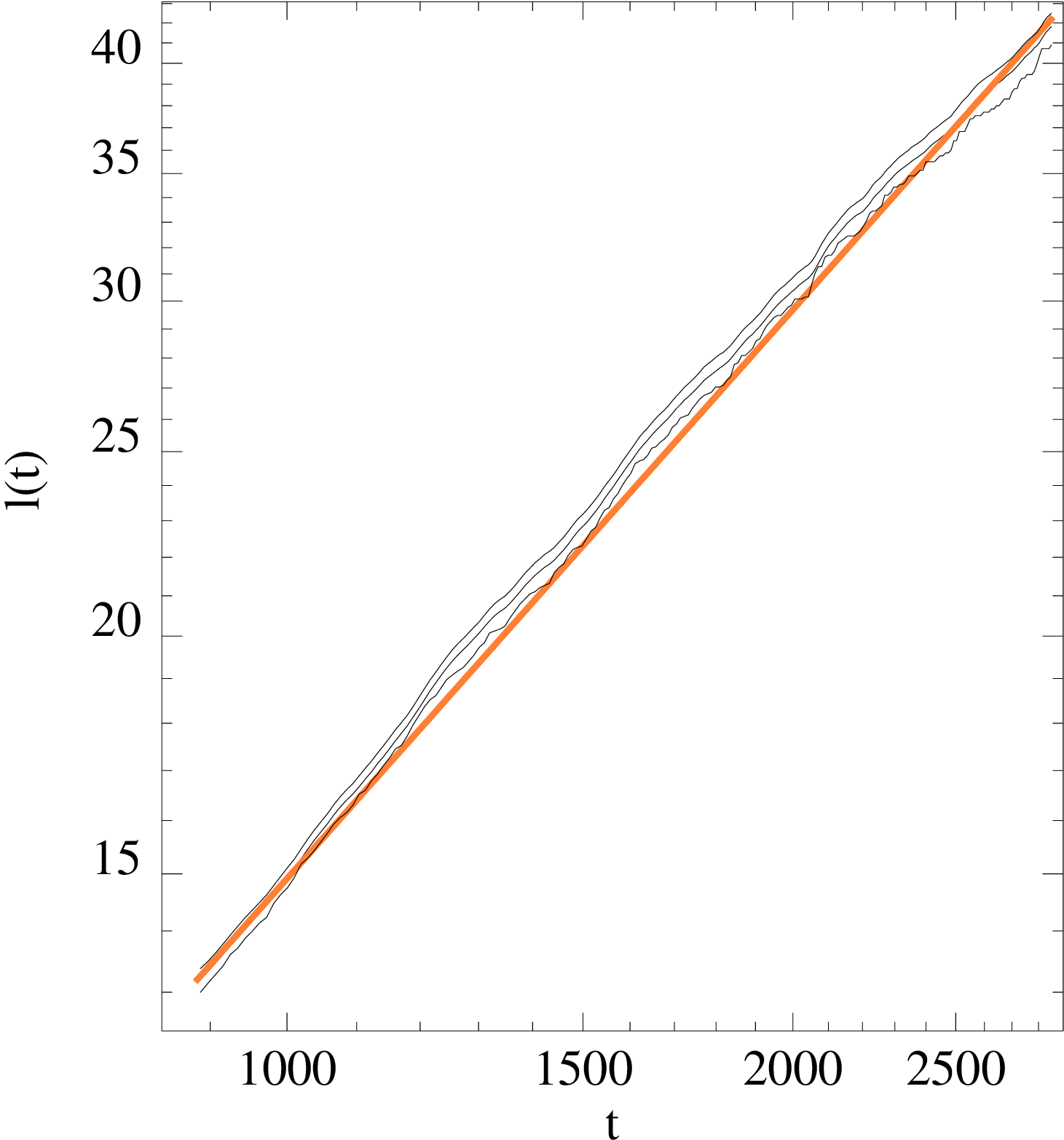,width=6cm}}
\end{minipage}
\begin{center}
(c)
\end{center}
\end{minipage}
\end{center}
\caption{(a) The droplet scaling state for $\eta=0.22$
(VH regime) for reduced times $t=1100,1500\mbox{
and } 2400$ respectively. (b) A section scaled by $l(t)$ is shown;
note the scale invariance.  (c) The evolution of $L^\#, L^\partial$ and
$L^\circ$ with time (where $L^\#< L^\partial <L^\circ $). All three length-scales
agree well with each other and approach $L\sim t$ (grey line).}
\label{fig:comp}
\end{figure}

This behavior is quite unexpected \cite{bhattacharya} and only partly
explicable. On one hand, while only local rearrangements (of order
parameter or interface) are needed to transform one to the other,
droplet and bicontinuous morphologies are very different: the largest
connected domains present are small in one case, and percolate in the
other. The issue is whether the `chain-reaction' of CIC is fast enough
to percolate the droplets \cite{nicolayev}: we find it is not fast
enough. On the other hand, it is just fast enough to keep going
indefinitely, whereas it might equally have stopped in finite time,
leaving classical, non-hydrodynamic coarsening processes to take
over. Note that under the CIC process, large domains can no longer
coarsen simply by deforming their tortuous boundaries, but have to
coalesce. This explains why scaling is recovered: all domains are
droplets and coarsen via the same mechanism.

In Fig.2 we see the counterpart of Fig.1 for a run that lies wholly in
IH. Here the initial droplet morphology (not shown) crosses over
rapidly to a bicontinuous scaling state seen previously for the IH
regime in two dimensions \cite{prl,furukawa}. The mechanism for the
rapid morphological transition is fluid inertia: two coalescing
droplets wobble violently as they fuse together, causing local as well
as distant coalescence events and rapid percolation of the initial
droplet structure. In the IH scaling state, the difference between
droplet or bicontinuous initial conditions has thus been lost and
local correlations inherited from early-time asymmetry do not have any
long-term effect. The observed coarsening mechanism, in which the
local topology is constantly being ruptured and reconnected by
underdamped interfacial oscillations, is somewhat reminiscent of the
`turbulent remixing' process hypothesized in \cite{grant} (see also
\cite{olvera}). (This is somewhat hard to ascertain from the still
pictures in Figure 2 but is easily seen in our animations
\cite{href}.) But in contrast to the prediction \cite{grant} that $l
\le O(t^{1/2})$, we see (as in \cite{furukawa}) a simple $l\sim
t^{2/3}$ scaling throughout the accessible IH regime.

Just as in $d=3$, we cannot rule out a further crossover to a regime
where the viscosity is so low that the growth rate is not limited by
the input of energy from the surface tension, so that the primary
balance in Eqs. \ref{ns},\ref{ad} is between inertial and viscous
effects. Kendon \cite{kendon2} considered this case in $d=3$ and
argued that $l\sim t^{2/3}$ might survive for the structural length
scale, so long as other lengths (describing the velocity field) do not
scale the same way. It would be interesting to pursue this suggestion
in $d=2$, although the  properties of $d=2$ and $d=3$
turbulence \cite{turbulence} are very different.

To establish scaling with confidence, Figure 1(c) and 2(c) are not
sufficient; we must combine several simulations with different
viscosities to span many decades on the $l(t)$ curve \cite{kendon}. As
shown in Fig.3, this yields very good data collapse onto a universal
scaling curve for all three length measures. A prefactor deviation
appears between $L^\#$ and the others in the IH regime, but this
results from the change in morphology (Fig.2), not a violation of
scaling. The asymptotes are (for lengths defined via $L^\circ$ and
$L^\partial$) $l=0.02 t$ and $l= 0.9 t^{2/3}$. These cross at $t= t^*
\simeq 2\times 10^5$, a value even larger than the $d=3$ case, whose
asymptotes (using $L = 2\pi \int S(k) dk/\int k S(k) dk$) are also
shown for comparison.

\begin{figure}
\begin{center}
\begin{minipage}{8cm}
\begin{minipage}{3.5cm}
\centerline{\psfig{figure= 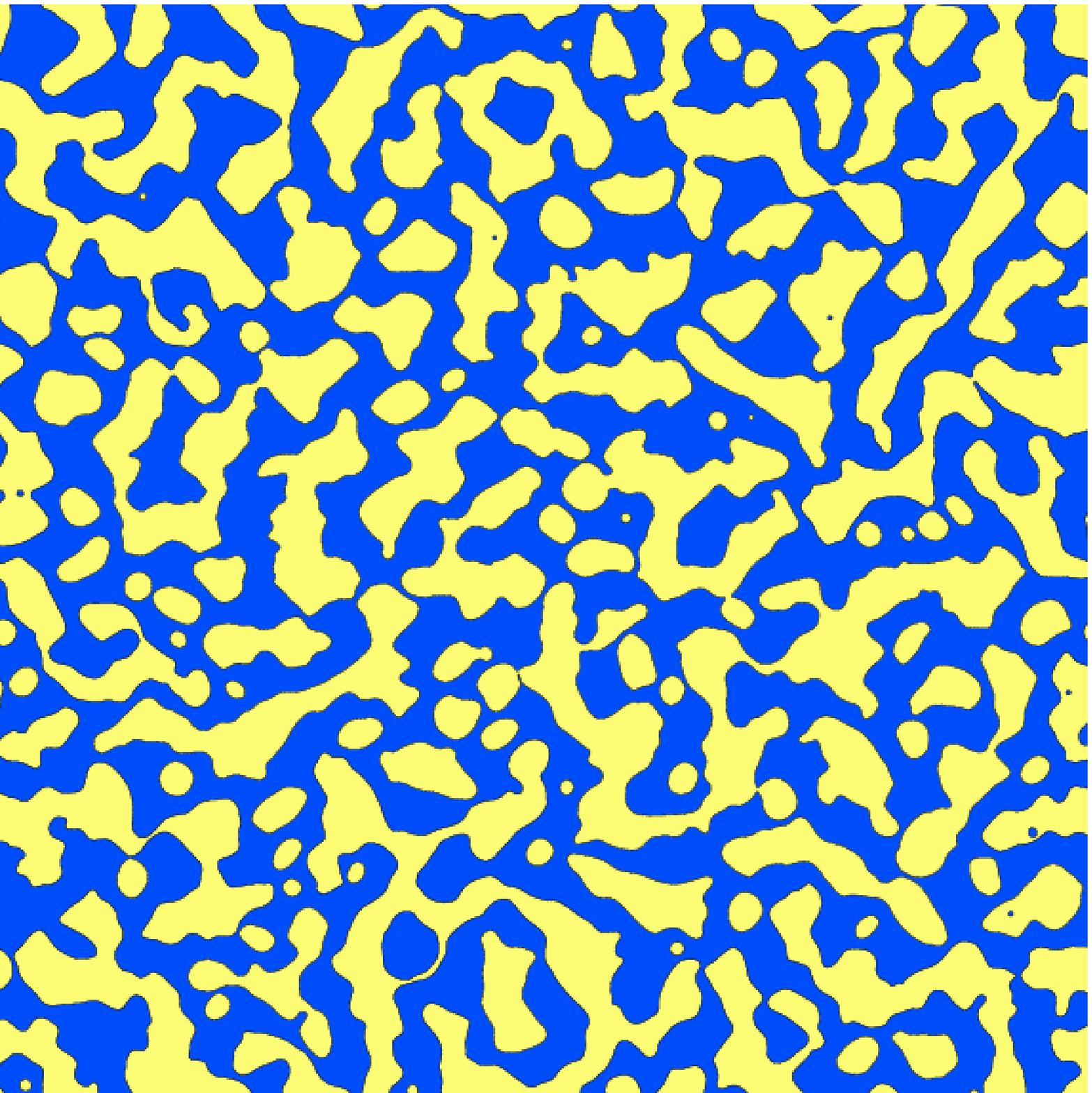,width=3cm}}
\end{minipage}
\hspace{0.3cm}
\begin{minipage}{3.5cm}
\centerline{\psfig{figure=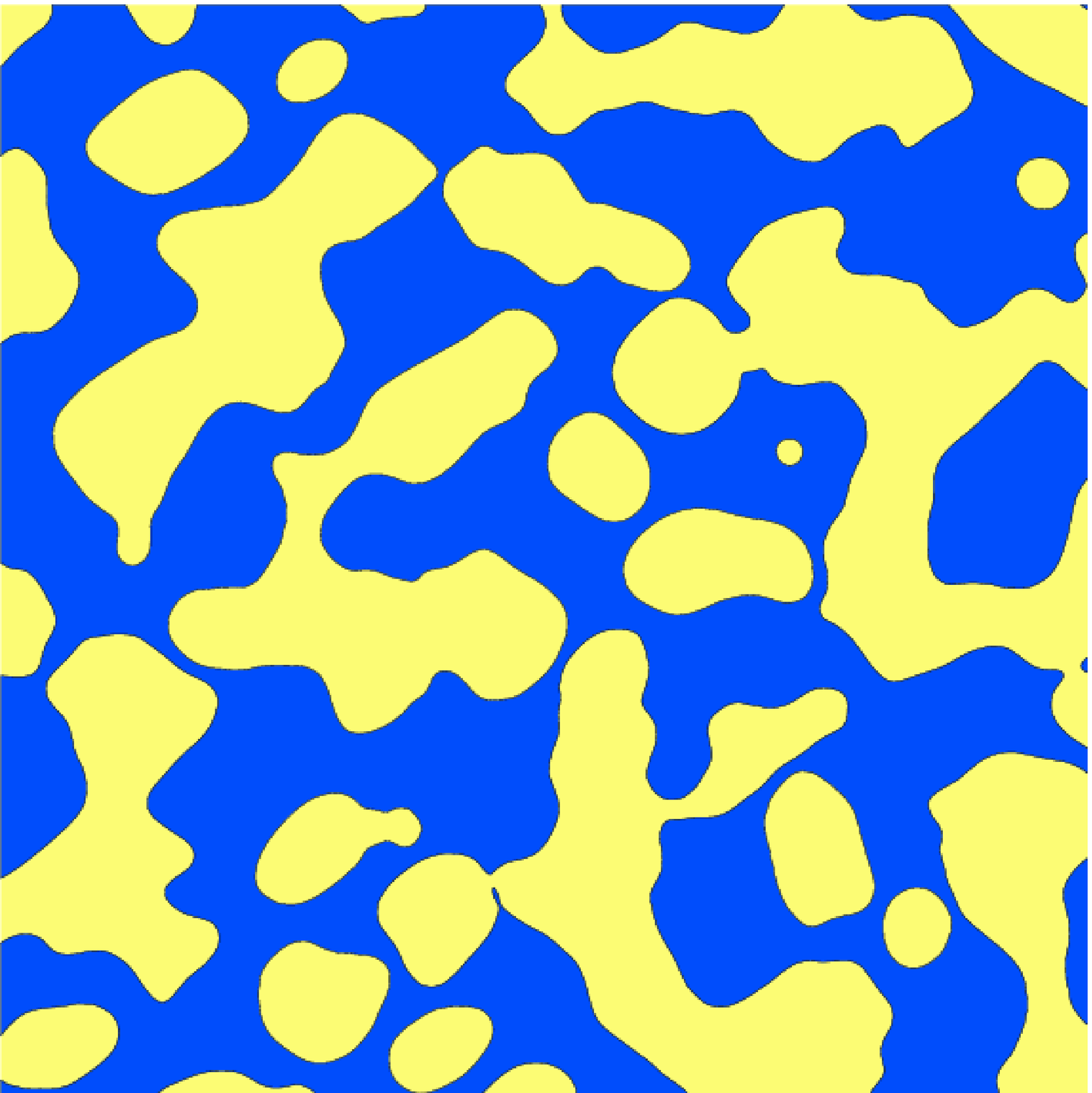,width=3cm}}
\end{minipage}\\ \vspace{0.1cm}

\begin{minipage}{3.5cm}
\centerline{\psfig{figure=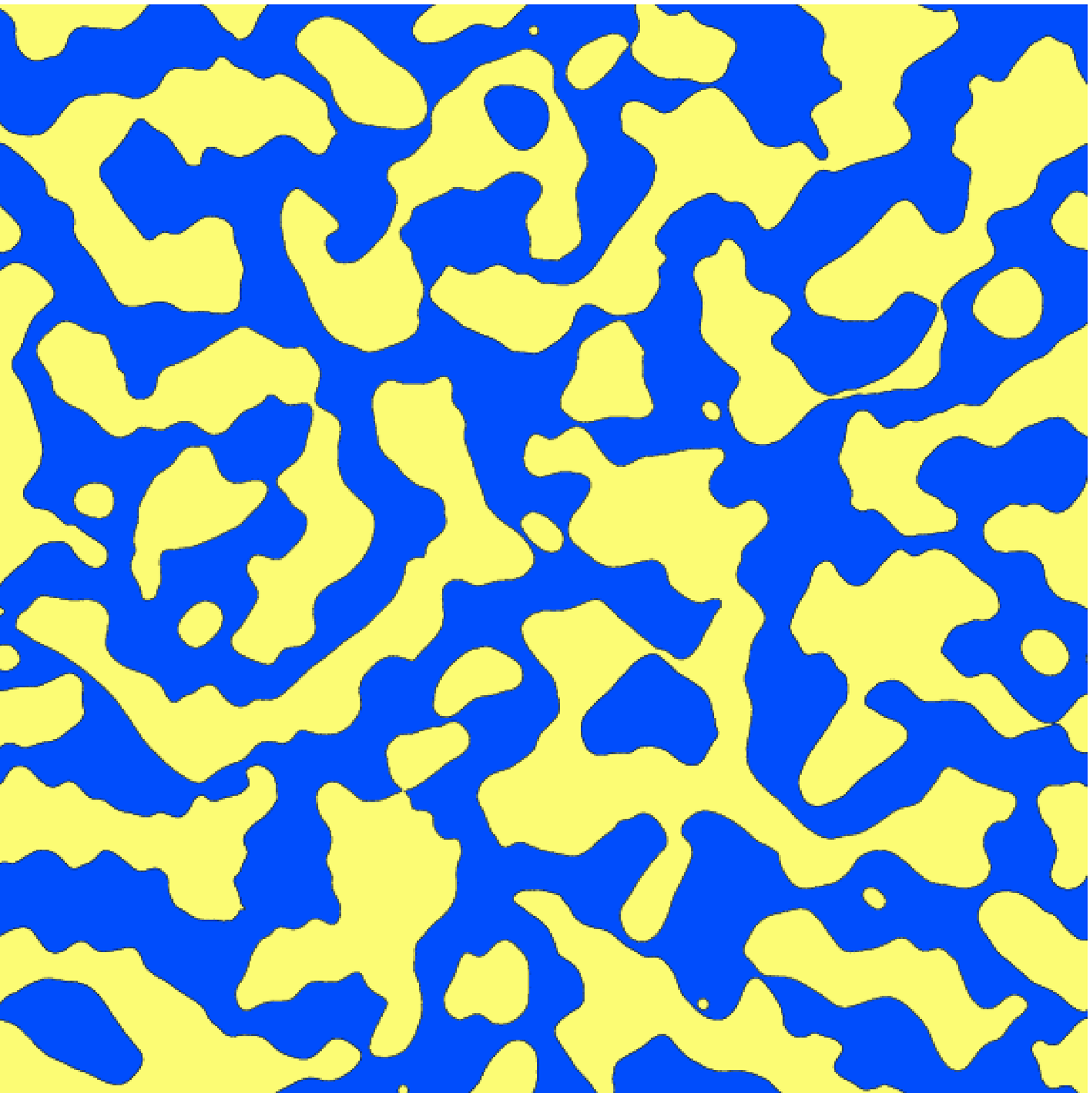,width=3cm}}
\end{minipage}
\hspace{0.3cm}
\begin{minipage}{3.5cm}
\centerline{\psfig{figure=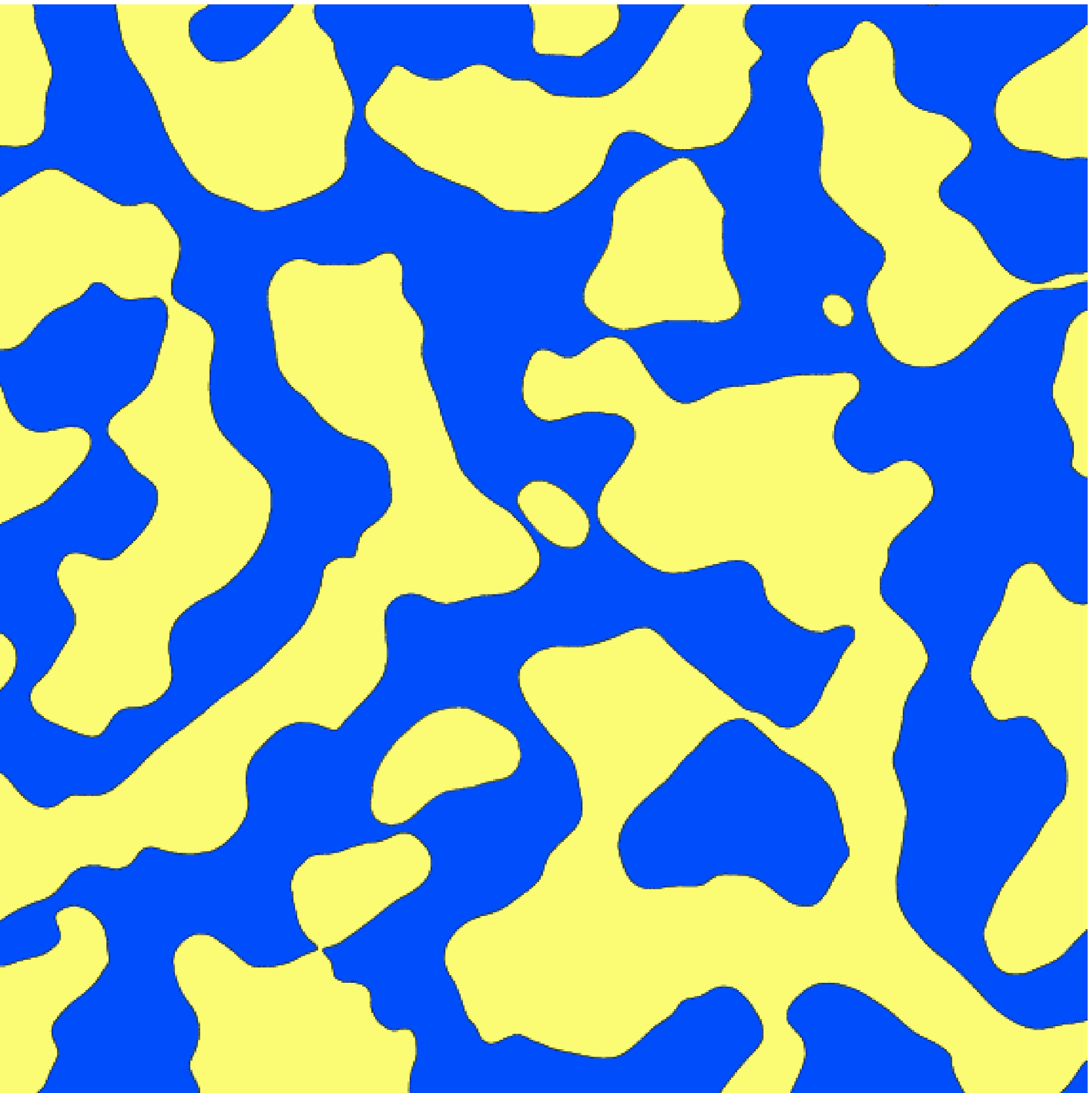,width=3cm}}
\end{minipage}\\ \vspace{0.1cm}

\begin{minipage}{3.5cm}
\centerline{\psfig{figure= 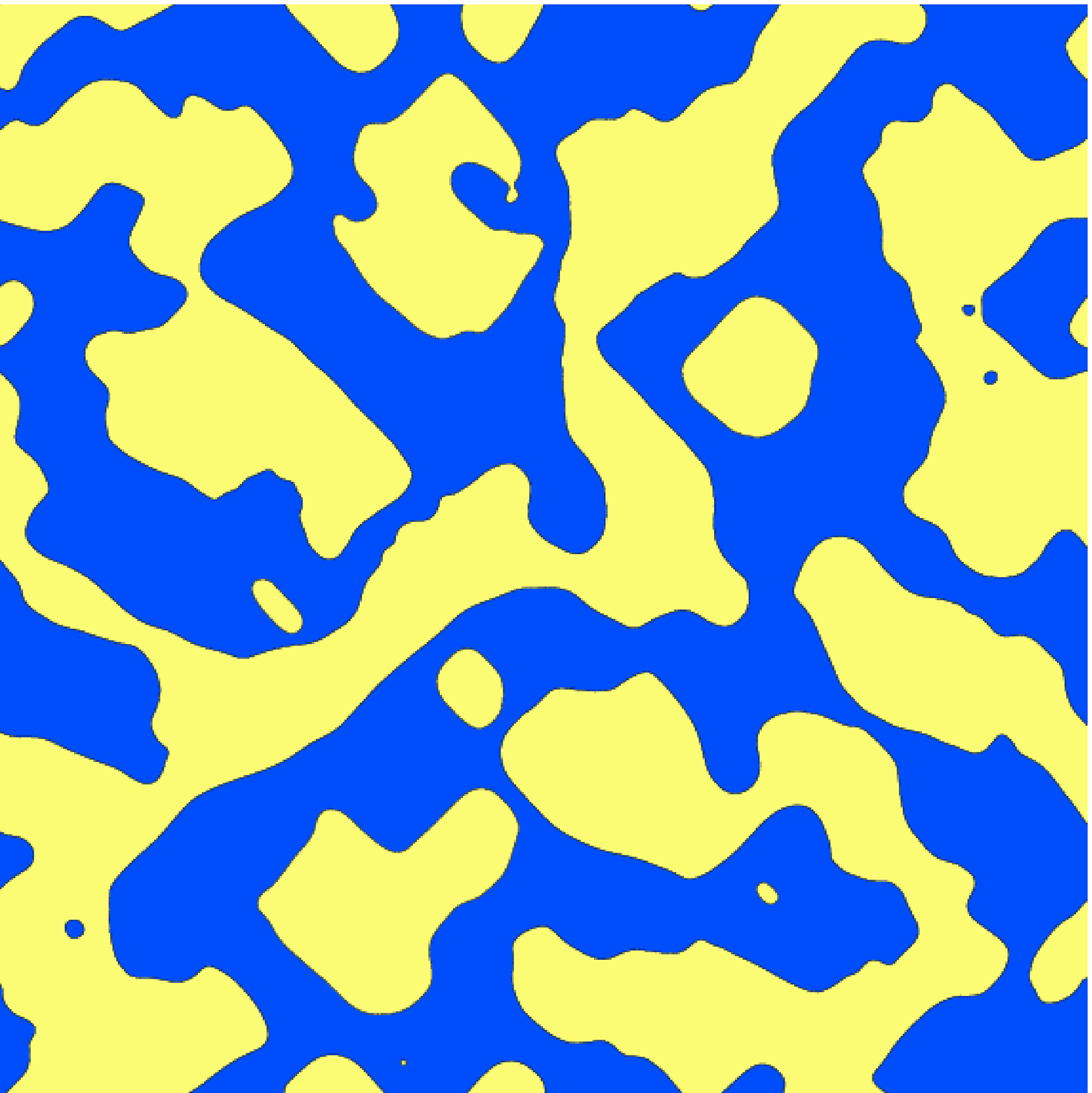,width=3cm}}
\end{minipage}
\hspace{0.3cm}
\begin{minipage}{3.5cm}
\centerline{\psfig{figure= snu0.0022_38436.eps,width=3cm}}
\end{minipage}
\begin{center}
(a) \hspace{2.5cm} (b) 
\end{center}
\begin{minipage}{8cm}
\centerline{\psfig{figure=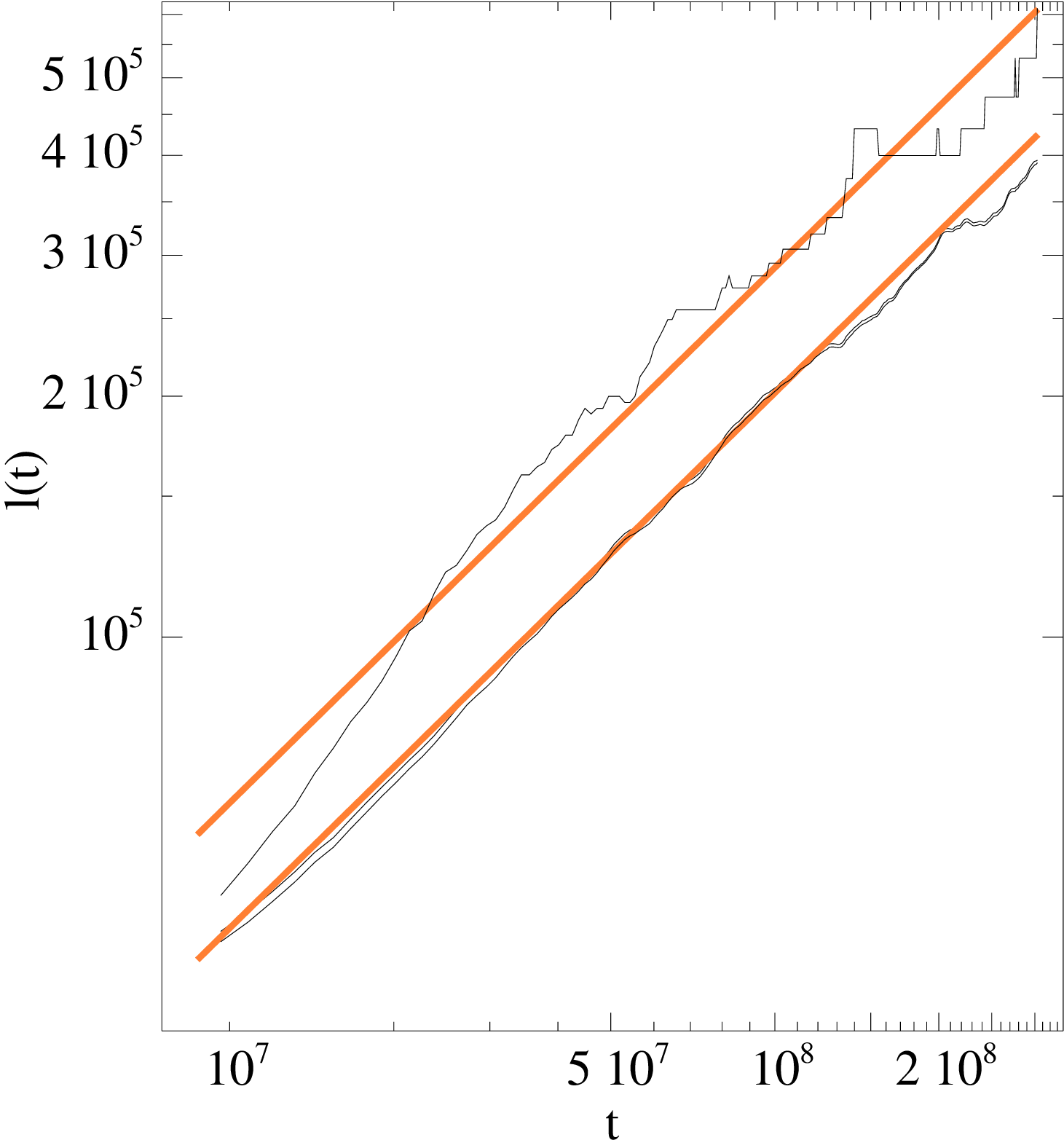,width=6cm}}
\end{minipage}
\begin{center}
(c)
\end{center}
\end{minipage}
\end{center}
\caption{Recovery of bicontinuous scaling state for $\eta= 0.0044$ (IH
regime). (a) The symmetric scaling state that rapidly evolved from the
initial droplet morphology for $t=2\times 10^7,\; 4.2\times
10^7\mbox{ and }9.3\times 10^7$. (b) Images scaled by $l(t)$, showing
dynamic scaling. (c) $L^\circ$,$L^\partial$ and $L^\#$ all scale as
$t^{2/3}$ but $L^\#$ is now larger that the other two measures,
indicating the change in morphology.}
\label{fig:comp2}
\end{figure}

In comparable units, our VH droplet growth is significantly
slower than for the $d=3$ bicontinuous VH scaling morphology. This
is reasonable: the amount of interface actually driving the coarsening
is much smaller since only droplets undergoing coalescence contribute
to the flow at any time, while in the bicontinuous case the entire
interface is in motion. Once the IH regime is entered the $d=2,3$
structures are both bicontinuous and the coefficients of $t^{2/3}$
become much more similar.

\begin{figure}
\centerline{\psfig{figure=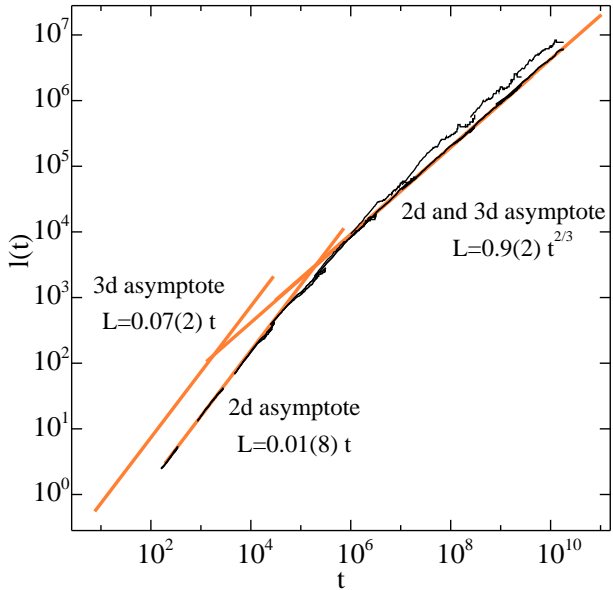,width=8cm}}
\caption{ Universal scaling curve $l(t)$ for the evolving droplet
state. Note the separation of $L^\#$ (on top) from $L^\circ$ and
$L^\partial$ at the crossover to the IH regime, indicating the change
in morphology. Asymptotes in grey; those for bicontinuous $d=3$
coarsening \protect\cite{kendon} are also shown.}
\end{figure}

To summarize, in this letter we have found an unexpected droplet
scaling state in the phase-ordering of two-dimensional symmetric
binary mixtures at 50/50 compositions. This implies that the local
order parameter correlations brought about by asymmetric early-time
dynamics ({\it e.g.} diffusion or viscoelasticity) can continue to
control the behavior even when the coarsening length has become very
large compared to the original length scale of these correlations, and
long after the asymmetry in the evolution equations has become
negligible.  The novel droplet scaling state coarsens with $l\approx
0.02 t$ in our reduced physical units; this linear scaling is
characteristic of the VH regime as in $d=3$, but with a reduced
prefactor. In the inertial regime the system crosses over to a full
symmetric morphology with a scaling law of $l\approx 0.9\;t^{2/3}$,
similar to the $d=3$ IH scaling state \cite{kendon}. The ultimate fate
of the IH regime in both $d=2$ and $d=3$ remains uncertain
\cite{grant,kendon2}.

Finally, we also performed some preliminary simulations of droplet
morphologies for symmetric $d=3$ fluids. As predicted in
\cite{nicolayev}, these morphologies always crossed over rapidly to a
bicontinuous morphology, even when initialized in the VH regime: this
contrasts strongly with our $d=2$ results. Thus it remains possible
that for $d=2$ the absence of scaling of the bicontinuous morphology,
and its restoration for a droplet state, is connected with the fact
that the 50/50 bicontinuous state is (by symmetry) precisely at the
percolation threshold. It would be interesting to see whether $d=3$
systems close to percolation show similar anomalies, and whether a
self-sustaining CIC droplet scaling is again recovered for
compositions in that region.

This work was funded in part by grant EPSRC (UK) GR/M56234. We thank
I. Pagonabarraga for discussions.

\def\jour#1#2#3#4{{#1} {\bf #2}, #3 (#4)}.
\def\tbp#1{{\em #1}, to be published}.
\def\inpr#1{{\em #1}, in preparation}.
\def\tit#1#2#3#4#5{{#1} {\bf #2}, #3 (#4)}

\def\ap{Adv. Phys.}
\def\arf{Ann. Rev. Fluid Mech.}
\def\epl{Euro. Phys. Lett.}
\def\ijmp{Int. J. Mod. Phys. C}
\def\jcp{J. Chem. Phys.}
\def\jpc{J. Phys. C}
\def\jpcs{J. Phys. Chem. Solids}
\def\jpco{J. Phys. Cond. Mat}
\def\jsp{J. Stat. Phys.}
\def\jf{J. Fluids}
\def\jfm{J. Fluid Mech.}
\def\jnnfm{J. Non-Newtonian Fluid Mech.}
\def\pfa{Phys. Fluids A}
\def\prl{Phys. Rev. Lett.}
\def\pr{Phys. Rev.}
\def\pra{Phys. Rev. A}
\def\prb{Phys. Rev. B}
\def\pre{Phys. Rev. E}
\def\pa{Physica A}
\def\pla{Phys. Lett. A}
\def\ps{Physica Scripta}
\def\roy{Proc. Roy. Soc.}
\def\rmp{Rev. Mod. Phys.}
\def\zpb{Z. Phys. B}


\begin{thebibliography}{200}
\bibitem{footeuler} Of course, these two morphologies have different
mean density of the Euler invariant, and in that sense, as well as in
the sense of percolation, have
`infinite range correlations'. Unlike correlations of the order
parameter $\phi$, however, these can evolve from a white-noise initial
state, by a purely local early-time dynamics in
finite time.
\end{thebibliography}
\end{document}